%% file: main.tex
\documentclass[sigconf]{acmart}

\AtBeginDocument{%
  }

\setcopyright{acmlicensed}
\copyrightyear{2025}
\acmYear{2026}
\acmDOI{XXXXXXX.XXXXXXX}
\acmConference[Conference acronym 'XX]{Make sure to enter the correct
  conference title from your rights confirmation email}{June 03--05,
  2018}{Woodstock, NY}
\acmISBN{978-1-4503-XXXX-X/2018/06}

\usepackage{wasysym}
\usepackage{mathtools}
\usepackage{lipsum}
\usepackage{graphicx}
\usepackage{caption, subcaption}
\usepackage{pifont}
\usepackage{ragged2e}

\usepackage{url}
\usepackage{amsmath, amsfonts, amsthm, bm}
\usepackage{soul}
\usepackage{tabularx, multirow}
\usepackage[ruled,vlined]{algorithm2e}
\usepackage{xcolor, colortbl}
\usepackage{xspace}
\usepackage{makecell}
\usepackage{enumitem}
\usepackage{booktabs}
\usepackage{array}
\usepackage{adjustbox}

\usepackage{pgfplots}
\pgfplotsset{compat=1.18}  

\usepackage{listings}
\input{dfn}

\begin{document}

\title{Offline Reasoning for Efficient Recommendation: LLM-Empowered Persona-Profiled Item Indexing}

\author{Deogyong Kim}
\authornote{Both authors contributed equally to this research.}
\affiliation{%
  \institution{Yonsei University}
  \city{Seoul}
  \country{South Korea}
  }
\email{legenduck@yonsei.ac.kr}

\author{Junseong Lee}
\authornotemark[1]
\affiliation{%
  \institution{Yonsei University}
  \city{Seoul}
  \country{South Korea}
  }
\email{brulee@yonsei.ac.kr}

\author{Jeongeun Lee}
\affiliation{%
  \institution{Yonsei University}
  \city{Seoul}
  \country{South Korea}
  }
\email{ljeadec31@yonsei.ac.kr}

\author{Dongha Lee}
\authornote{Corresponding author}
\affiliation{%
    \institution{Yonsei University}
    \city{Seoul}
    \country{South Korea}
    }
\email{donalee@yonsei.ac.kr}

\author{Changhoe Kim}
\affiliation{%
  \institution{NAVER}
  \city{Seongnam}
  \country{South Korea}
  }
\email{andres.chkim@navercorp.com}

\author{Junguel Lee}
\affiliation{%
  \institution{NAVER}
  \city{Seongnam}
  \country{South Korea}
  }
\email{junguel.lee@navercorp.com}

\author{Jungseok Lee}
\affiliation{%
  \institution{NAVER}
  \city{Seongnam}
  \country{South Korea}
  }
\email{jungseok.lee@navercorp.com}

\renewcommand{\shortauthors}{Trovato et al.}

\begin{abstract}
\input{TEX/000abstract}
\end{abstract}


\maketitle

\vspace{-5pt}
\section{Introduction}
\label{sec:intro}
\input{TEX/010introduction_WWW}

\section{Related Work}
\label{sec:relwork}
\input{TEX/020relatedwork_WWW}
\section{Proposed Method}
\label{sec:method}
\input{TEX/030method_WWW}

\section{Experiments}
\label{sec:exp}
\input{TEX/040experiments_WWW}

\section{Conclusion}
\label{sec:conclusion}
\input{TEX/050conclusion_WWW}


\bibliographystyle{ACM-Reference-Format}
\bibliography{BIB/refs}


\end{document}

%% file: dfn.tex
\definecolor{DarkGreen}{RGB}{30,130,30}
\newcommand{\cmark}{\textcolor{DarkGreen}{\ding{51}}}
\newcommand{\xmark}{\textcolor{red}{\ding{55}}}%

\definecolor{DarkYellow}{RGB}{255,204,0}

\newcommand{\amazon}{Amazon-Books\xspace}
\newcommand{\yelp}{Yelp\xspace}

\newcommand{\bprmf}{BPR-MF\xspace}

\newcommand{\lightgcn}{LightGCN\xspace}

\newcommand{\rankvicuna}{RankVicuna\xspace}
\newcommand{\tallrec}{TALLRec\xspace}

\newcommand{\xrec}{XRec\xspace}

\newcommand{\expert}{EXP3RT\xspace}

\newcommand{\zsllm}{ZS-LLM\xspace}

\newcommand{\proposed}{\textsc{Persona4Rec}\xspace}

\newcommand{\smallsection}[1]{{\vspace{0.05in} \noindent \bf {#1.\hspace{5pt}}}}

%% file: TEX/000abstract.tex
Recent advances in large language models (LLMs) offer new opportunities for recommender systems by capturing the nuanced semantics of user interests and item characteristics through rich semantic understanding and contextual reasoning.
In particular, LLMs have been employed as rerankers that reorder candidate items based on inferred user–item relevance. 
However, these approaches often require expensive online inference-time reasoning, leading to high latency that hampers real-world deployment.

In this work, we introduce \proposed, a recommendation framework that performs offline reasoning to construct interpretable \textit{persona} representations of items, enabling lightweight and scalable real-time inference.
In the offline stage, \proposed leverages LLMs to reason over item reviews, inferring diverse user motivations that explain why different types of users may engage with an item; these inferred motivations are materialized as persona representations, providing multiple, human-interpretable views of each item.
Unlike conventional approaches that rely on a single item representation, \proposed learns to align user profiles with the most plausible item-side persona through a dedicated encoder, effectively transforming user–item relevance into user–persona relevance.
At the online stage, this persona-profiled item index allows fast relevance computation without invoking expensive LLM reasoning. Extensive experiments show that \proposed achieves performance comparable to recent LLM-based rerankers while substantially reducing inference time.
Moreover, qualitative analysis confirms that persona representations not only drive efficient scoring but also provide intuitive, review-grounded explanations. 
These results demonstrate that \proposed offers a practical and interpretable solution for next-generation recommender systems.\footnote{https://github.com/legenduck/PERSONA4REC}

%% file: TEX/010introduction_WWW.tex
\begin{figure}[!t]
    \centering
    \includegraphics[width=\linewidth]{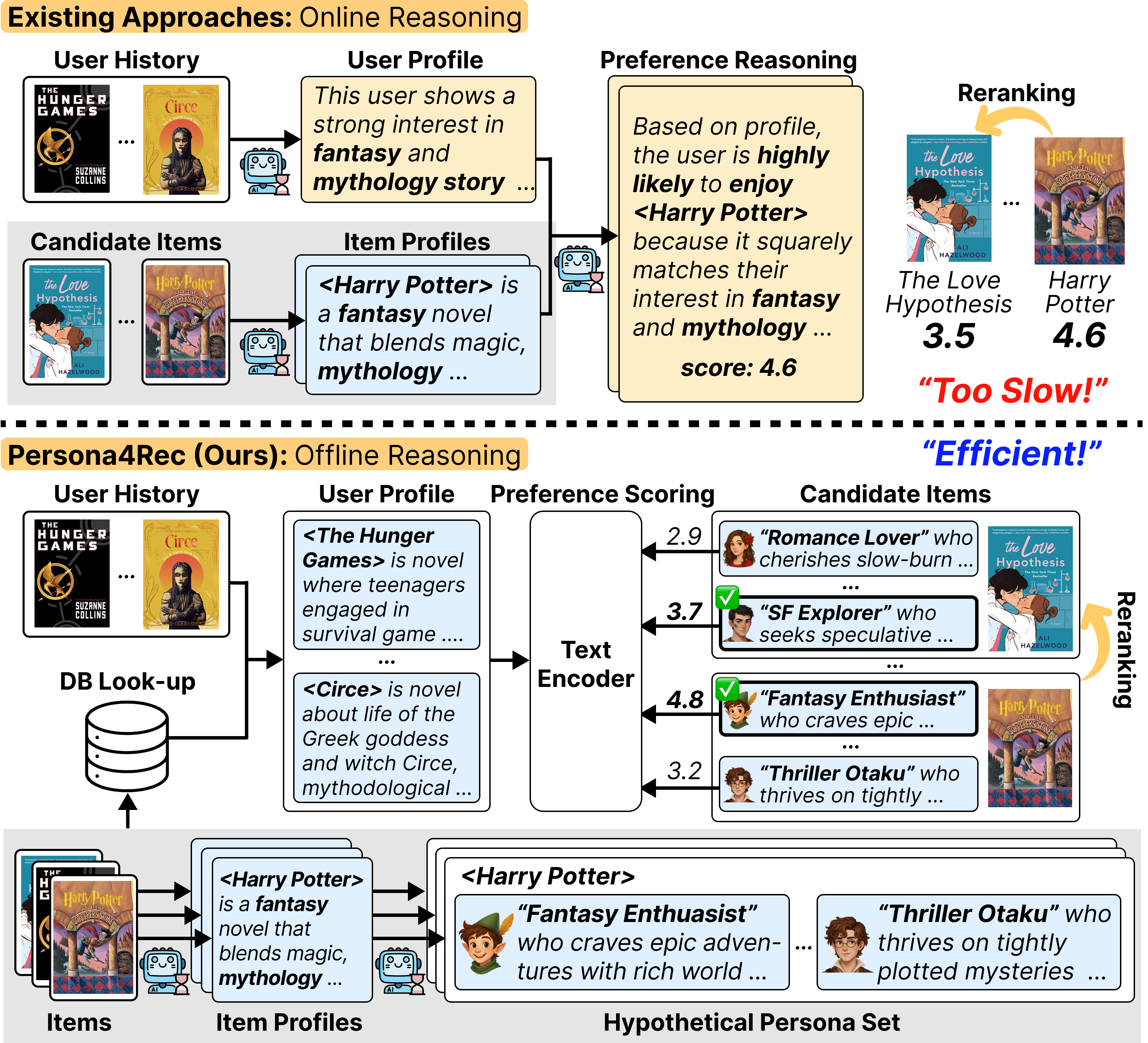}
    \caption{Comparison between existing LLM-based item rerankers (Upper) and our \proposed (Lower). \proposed shifts LLM reasoning from online inference to offline persona construction, enabling real-time recommendation via lightweight similarity scoring.}
    \label{fig:intro}
    \vspace{-5pt}
\end{figure} 

Traditional recommender systems (RSs) have achieved remarkable success with collaborative filtering (CF)~\cite{he2017neural,wang2019neural,he2020lightgcn, wu2021self, lee2024graph} and deep learning~\cite{zhang2019deep, sun2019bert4rec, li2023deep}, but they largely focus on interaction signals and limited metadata, which often fail to capture the rich semantics behind user preferences and item characteristics~\cite{geng2022recommendation, li2023text,zhang2025collm}. 
Recent advances in large language models (LLMs) have opened new opportunities for recommendation by leveraging their strong semantic understanding and contextual reasoning capabilities~\cite{zhao2024recommender,lee2025hippo,heo2025can,lee2025imagine}.
These abilities allow LLMs to process complex textual inputs such as reviews and item descriptions, leading to richer user and item representations and more personalized recommendations~\cite{liu2025large, zhang2025llm}.

One widely adopted approach to employing LLMs in recommendation is to construct user profiles~\cite{ren2024representation, purificato2024user, kronhardt2024personaer}.
These profiles are inferred from user interaction histories, often by leveraging metadata and reviews, to capture their overall tendencies across behavioral patterns.
By providing a compact yet expressive representation of user interests, user profiles enable models to enhance personalization~\cite{yang2023palr, zhang2024guided} and serve as a foundation for various downstream recommendation tasks.
For instance, they support explainable recommendation by aligning user interests with item attributes to generate concise, personalized justifications~\cite{wozniak2025improving}, improve ranking quality by refining candidate ordering~\cite{kim2025driven}, and simulate user behavior in dynamic recommendation scenarios~\cite{cai2025large}.

Another growing line of work has explored LLMs as rerankers to refine the candidate items retrieved by first-stage recommenders (i.e., candidate generators) such as CF models~\cite{he2020lightgcn}.
Early studies adopt prompt-only, zero/few-shot listwise reranking~\cite{sun2023chatgpt,ma2023zero,pradeep2023rankvicuna,pradeep2023rankzephyr, hou2024large, adeyemi2024zero}, where LLMs are prompted to directly output a permutation of the candidate set. 
Following approaches introduce domain adaptation and fine-tuning techniques~\cite{bao2023tallrec,yue2023llamarec, luo2025recranker} to improve controllability and stability for recommendation tasks.
More recent methods leverage LLM reasoning to infer user preferences from raw signals such as interaction histories and reviews~\cite{kim2025driven,gao2025llm4rerank,besta2024graph}, while others adopt reinforcement learning to further align reranking with recommendation objectives~\cite{sun2024rlrf4rec, lu2024aligning}.
Overall, this growing line of research highlights how LLMs can fundamentally advance recommendation by moving beyond traditional interaction signals toward more semantically rich and context-aware modeling.

Despite these advances, such approaches remain impractical for real-world deployment, where real-time recommendation is essential for user experience.
Their online (inference) stage 
involves multiple time-consuming reasoning steps that cause substantial latency, whereas computation time is far less critical in the offline (training) stage (Figure~\ref{fig:intro}, Upper).
Specifically, it includes (1) constructing user profiles from historical interactions such as ratings and reviews, (2) generating item profiles from metadata and user feedback, and (3) performing reasoning-based reranking by comparing candidate items through these profiles.
Because user histories evolve dynamically, RSs must repeatedly infer user profiles from the latest interactions and then reason over them to assess item relevance, which substantially increases inference latency and poses serious challenges for efficiency and scalability.


To address these challenges, we introduce a novel approach that achieves low-latency inference by leveraging \textit{item-side reasoning pre-computed offline}, thereby supporting both the efficient construction of user profiles and the efficient scoring of candidate items (Figure~\ref{fig:intro}, Lower).
This offline reasoning step extracts fine-grained, subjective preference aspects from item reviews, which can later be used to dynamically compose user profiles from their interaction histories. 
These extracted aspects are then re-organized into an {item-side knowledge index} that enables recommender systems to efficiently compute user–item relevance at inference time through lightweight semantic similarity scoring.
This design not only enables scalable, real-time recommendation but also yields interpretable outputs, as each recommendation is grounded in review-based reasoning derived from the offline process.


In this work, we propose \proposed, which introduces the notion of \textit{hypothetical user personas}---interpretable profiles that encode plausible rationales, extracted from item reviews, for why certain users might engage with the item.
Rather than relying on a single representation per item, \proposed represents items through multiple personas to reflect potential user motivations in a human-understandable form;
each persona can also be interpreted as a latent \textit{user segment} likely to prefer the corresponding item.
During inference, user profiles aggregated from historical interactions are aligned with the most relevant persona via a lightweight encoder, trained on interaction-derived user-persona pairs to reflect realistic engagement patterns;
this replaces direct user-item matching with user–persona alignment. 
In this way, recommendations are generated through efficient similarity scoring over a \textit{persona-profiled item index}, where each persona serves as an independent retrieval unit linked to its source item,
and the selected persona naturally provides an intuitive rationale for each ranked item.

Our extensive experiments mainly investigate the trade-off between recommendation accuracy and inference efficiency, comparing \proposed against state-of-the-art LLM-based rerankers.
The results show that \proposed 
achieves comparable accuracy to state-of-the-art LLM-based rerankers while reducing inference time by up to 99.6\%, demonstrating its practicality for large-scale deployment.
Moreover, qualitative analysis confirms that the selected personas provide user-grounded and interpretable explanations.

The main contributions of this work are summarized as follows:
\begin{itemize}
    \item We propose \proposed, a novel framework that shifts reasoning offline and enables efficient-yet-effective reranking through lightweight user–item relevance scoring.
    \item Extensive experiments demonstrate that \proposed ach-
    ieves real-time efficiency while maintaining performance comparable to state-of-the-art LLM-based rerankers.
    \item By leveraging precomputed item personas, \proposed naturally provides faithful and human-readable explanations for recommendation outcomes.
\end{itemize}

%% file: TEX/020relatedwork_WWW.tex

\begin{figure*}[t]
    \centering
    \includegraphics[width=\textwidth]{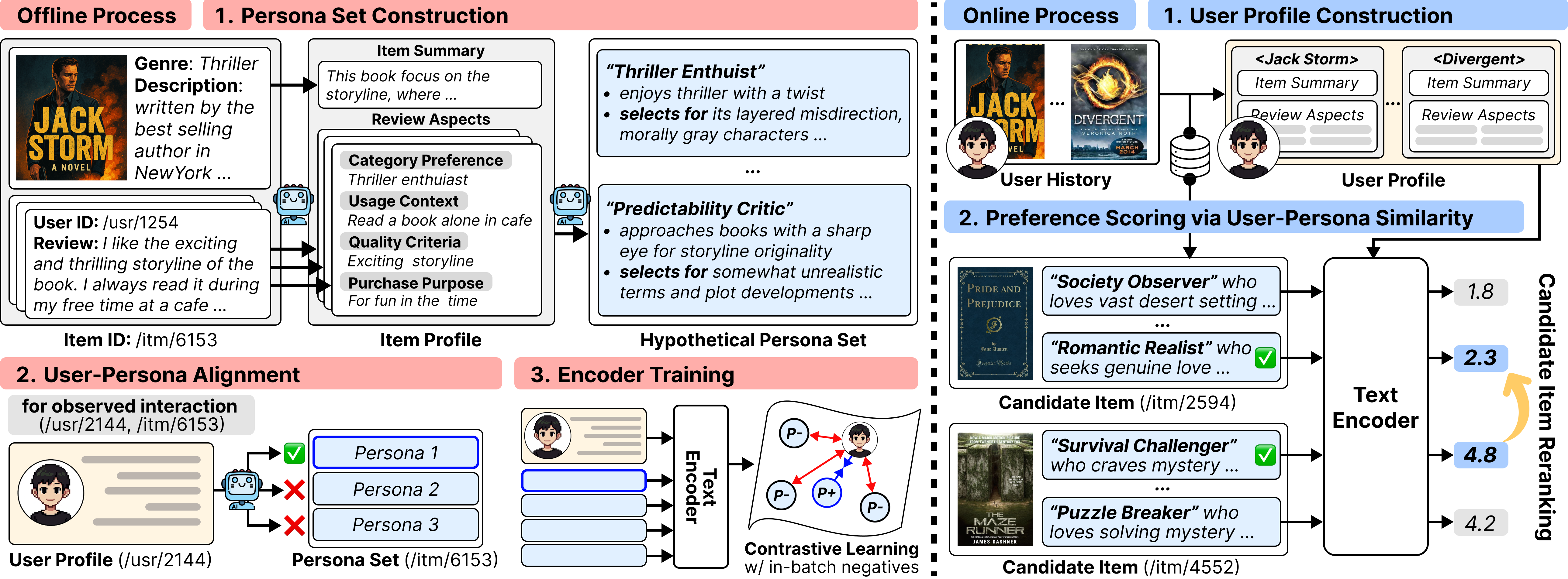}
    \vspace{-0.4cm}
    \caption{Overview of our \proposed framework. The offline process (Left) constructs item personas and trains a user-persona alignment encoder. The online process (Right) reranks candidates via efficient similarity scoring}
    \label{fig:overview}
\end{figure*}

\input{TBL/011_Compare_Previous_method}

\subsection{LLM-based Reranking in Recommendation}
\label{subsec:rw_llm_rerank}
Early studies demonstrated that LLMs guided by constructed prompts can act as rerankers without task-specific training, often in a {listwise} form that outputs a permutation over a candidate set~\cite{sun2023chatgpt,ma2023zero,pradeep2023rankvicuna,pradeep2023rankzephyr,tamber2023scaling}.
To alleviate long-context costs, follow-up work explored {pairwise} prompting via PRP~\cite{qin2023large} and {setwise} prompting that compares small subsets~\cite{zhuang2024setwise}. 
While these approaches report strong accuracy, their reliance on autoregressive decoding and long contexts often results in high latency and forces truncation or sliding-window heuristics in practice~\cite{sun2023chatgpt,tamber2023scaling}.

Recent approaches aligns LLMs with recommendation objectives through instruction tuning or reinforcement learning. 
TALLRec~\cite{bao2023tallrec} aligns LLMs with recommendation tasks by instruction tuning on instruction–response pairs, enabling more accurate user–item ranking, and RLRF4Rec~\cite{sun2024rlrf4rec} directly optimizes a reranker from recommender feedback. 
Review-driven methods such as EXP3RT~\cite{kim2025driven} extract preference evidence from reviews to improve rating prediction and top-$k$ reranking. 
LLM4Rerank~\cite{gao2025llm4rerank} further integrates multiple criteria---including accuracy, diversity, and fairness---by modeling them as interconnected nodes and applying CoT-style reasoning to automatically navigate these nodes during the reranking process.
Despite these advances, existing reranker designs still depend on costly inference-time reasoning to extract or align preferences, which limits their scalability in real-time recommendation.

\vspace{-5pt}
\subsection{User Profiling for Recommendation}
\label{subsec:rw_profiles}

Early recommender systems relied on \textit{static or schema-bound} user profiles, such as demographics or genre preferences~\cite{chen2007content}. 
While these profiles offered a simple way to represent users, they lacked the capacity to capture nuanced or context-dependent preferences.  
Subsequent feature-engineered approaches, such as UPCSim~\cite{widiyaningtyas2021user}, attempted to measure profile similarity by correlating user and content attributes. 
However, hand-crafted features still struggled to reflect fine-grained signals and dynamic shifts in user interests.

To move beyond these limitations, recent studies employ LLMs to \textit{summarize, enrich, or generate} user profiles directly from interaction histories and side texts. 
 For example, RLMRec~\cite{ren2024representation} integrates LLM-guided representations into collaborative filtering (CF), while KAR~\cite{xi2024towards} and GPG~\cite{zhang2024guided} demonstrate that natural-language profile generation can enhance personalization.
Other works such as PALR~\cite{yang2023palr} and LettinGo~\cite{wang2025lettingo} further explore LLM-native profile modeling conditioned on user histories.
Taken together, existing work shows the strength of LLMs in modeling user preferences, but most approaches rely on online reasoning at inference time, which incurs high latency. In contrast, our work shifts both user- and item-side profiling to an offline stage, enabling efficient and scalable recommendation while retaining semantic richness.

%% file: TBL/011_Compare_Previous_method.tex

\begin{table}[t]
\centering
\caption{Comparison of text-based reranking methods in terms of efficiency and reasoning paradigm.
``Reasoning'' denotes when the reasoning process is performed (online during inference or offline as a pre-computation step).}
\vspace{-0.2cm}
\label{tab:related_comparison_resize}
\resizebox{0.99\linewidth}{!}{
\begin{tabular}{lcccc}
\toprule
\textbf{Method} & \textbf{Latency} & {\textbf{Scalability}} & \textbf{Reasoning} & \textbf{Explain.} \\
\midrule
\zsllm~\cite{hou2024large}        & High & Low & Online & \xmark \\
\rankvicuna~\cite{pradeep2023rankvicuna}    & High & Low & Online & \xmark \\
\tallrec~\cite{bao2023tallrec}       & Middle  & Low & Online & \xmark \\
\expert~\cite{kim2025driven}        & High & Low & Online & \cmark \\
\midrule
{\proposed (ours)} & \textbf{Low} & \textbf{High} & \textbf{Offline} & \cmark \\
\bottomrule
\end{tabular}
}
\end{table}

%% file: TEX/030method_WWW.tex

\label{sec:method}

The key idea of \proposed is to shift the costly reasoning process from online (inference) stage to the offline stage by precomputing review-grounded item personas. 
Specifically, during offline stage, \proposed derives multiple \textit{personas} for each item, where each persona captures a distinct motivation or preference pattern associated with the item.
These personas are organized into a \textit{persona-profiled item index} and serve as the primary units for aligning with user profiles, which summarize user historical interactions, rather than relying on direct alignment with raw item information.
During online 
stage, user profiles are efficiently matched against pre-indexed personas, enabling real-time recommendation without invoking expensive reasoning.
Moreover, the selected persona naturally provides a human-interpretable explanation for the recommendation, grounded in evidence from actual user reviews.
As illustrated in Figure~\ref{fig:overview}, \proposed consists of two stages:

\begin{itemize}[leftmargin=1.2em, itemsep=2pt, topsep=2pt]
    \item \textbf{Offline process:} 
    \proposed analyzes item information and reviews to generate multiple personas representing diverse user motivations. These personas are then paired with user profiles 
    to produce user–persona alignment signals, which are used to train a lightweight encoder that embeds both into a shared space and constructs a persona-profiled item index for efficient scoring.
    
    \item \textbf{Online process:} 
    During inference, \proposed constructs user profiles on-the-fly from recent interaction histories, leveraging the precomputed persona representations. 
    For each candidate item, its personas are scored against the user profile via efficient similarity computation, and relevance score is determined by the best-matching persona, which is used for reranking.
\end{itemize}

\subsection{Offline Reasoning}
\label{subsec:offline}

The offline 
stage of \proposed performs item-side reasoning through three coordinated steps that transform raw user reviews into structured personas, providing alignment signals for training the user–persona encoder:

\begin{enumerate}[leftmargin=1.2em,itemsep=2pt,topsep=2pt]
\item \textbf{Persona Construction (\S\ref{subsubsec:persona_construction}):} 
\proposed integrates item metadata and user reviews to construct multiple distinct and interpretable personas for each item, enabling fine-grained alignment with diverse user preferences.

\item \textbf{User Profile–Persona Matching (\S\ref{subsubsec:persona_matching}):} 
\proposed pairs each user profile with the most relevant persona of an interacted item using an LLM-as-a-judge paradigm, producing user–persona alignment signals.

\item \textbf{Encoder Training (\S\ref{subsubsec:encoder_training}):} 
\proposed leverages the resulting matched user–persona pairs to train a lightweight encoder that assigns higher similarity scores to matched pairs than to mismatched ones.
\end{enumerate} 
This structured reasoning pipeline leverages an off-the-shelf LLM $\mathcal{M}$
to perform profile 
construction and reasoning over reviews,\footnote{We use \texttt{gpt-4o-mini} for all LLM tasks in this paper.} 
transforming text into structured supervision for encoder training.

\subsubsection{Persona Construction}
\label{subsubsec:persona_construction}

This step transforms item-side information—objective metadata and subjective review signals—into multiple interpretable personas per item, each representing a distinct user motivation for engaging with the item.
Metadata provides factual context about what the item is, while reviews reveal why users engaged with it; integrating both enables the LLM $\mathcal{M}$ to infer latent user motivations and generate hypothetical user profiles that capture plausible intent beyond explicit review content.
~\footnote{Detailed prompts used for persona construction are provided in 
our code repository.}

\smallsection{Item Summary Generation (\textit{Objective} Information)}
For each item $i$, we instruct the LLM $\mathcal{M}$ to produce a concise summary $s_i$ that captures its core identity from the item metadata $m_i$, which include fields such as title, description and category.
\begin{equation}
    s_i = \mathcal{M}\big(m_i, \mathcal{I}_{\mathrm{sum}}\big).
\end{equation}
Here, $\mathcal{I}_{\mathrm{sum}}$ denotes an instruction prompt to summarize the information of the item.
The resulting summary provides a compact representation of the item and serves as an objective reference point for interpreting review-based signals.
Importantly, the summary also enables persona generation for items with sparse or missing reviews, allowing metadata alone to support effective persona construction in cold-start settings that commonly occur in practice.

\smallsection{Aspect Extraction (\textit{Subjective} Information)}  
For item $i$, the LLM $\mathcal{M}$ extracts an aspect tuple $a_{u,i}$ from each review by user $u$:
\begin{equation}
    a_{u,i} = \mathcal{M}(r_{u,i}, \mathcal{I}_{\mathrm{asp}})
\end{equation}
Here, $\mathcal{I}_{\mathrm{asp}}$ denotes an instruction prompt that specifies a domain-specific schema with slots such as \textit{category preference}, \textit{purchase purpose}, \textit{quality criteria}, and \textit{usage context}.
To ensure sufficient informational coverage, tuples with 
more than 75\% of slots being \texttt{null} fields are discarded.
The remaining tuples constitute the item-level aspect pool $\mathcal{A}_i$, which serves as input for the subsequent persona generation process.
Unlike metadata, which describes what the item is, these aspects capture how users actually experienced and valued the item—providing subjective signals essential for understanding diverse user motivations and informing persona construction.

\smallsection{Persona Generation}
Finally, the LLM $\mathcal{M}$ integrates the item summary $s_i$(objective context) and review-grounded aspect pool $\mathcal{A}_i$ (subjective review signals)to generate personas:
\begin{equation}
\mathcal{M}:(s_i, \mathcal{A}_i, \mathcal{I}_{\mathrm{per}}) \to \Pi_i = \{\pi_i^{(1)}, \ldots, \pi_i^{(K)}\},
\label{eq:persona_gen_revised}
\end{equation}
Here, $\mathcal{I}_{\mathrm{per}}$ denotes an instruction prompt for persona generation. The number of personas $K\in[2,7]$ varies based on the diversity of review signals: items with heterogeneous feedback yield more personas, while items with uniform reviews produce fewer. This adaptive range balances persona expressiveness against indexing and scoring overhead.
Each persona $\pi_i^{(k)}$ is a structured record of three 
fields:
\textit{name} simply describes the persona's core identity,
\textit{description} provides a brief summary outlining its general tendencies, motivations, and overall attitude toward the item, and
\textit{preference rationale} explains why this persona would appreciate the item, grounded in evidence from $\mathcal{A}_i$ such as review patterns or aspect-level cues.
These fields are serialized into a unified text template.

Table~\ref{tab:persona-example-dark} illustrates an example of how review aspects are aggregated into a coherent persona. 
For the item ``The Shadow in the Glass'', its review (written by user \textit{AEPTNCI3X5}) and the extracted aspects are encapsulated in the persona ``\textbf{P3: Dark Storyline Seeker}'',
which captures readers drawn to tension, moral ambiguity, and darker reinterpretations of classic tales.

\input{TBL/033_Persona_example}

\subsubsection{User-Persona Alignment}
\label{subsubsec:persona_matching}

This step pairs user profiles with the most relevant persona from each interacted item via LLM-based alignment, producing supervision signals for encoder training.

First, we build user profile $P_u$, given a user’s interaction history $H_u$, that summarizes the user’s preferences across recently consumed items by combining objective and subjective signals:
\begin{equation}
P_u = \{ (s_i, a_{u,i}) \mid i \in H_u \}
\label{eq:user_profile}
\end{equation}
where we collect, for each item $i$ in the user's history, the item summary $s_i$ paired with the aspect tuple $a_{u,i}$ extracted from the user's review $r_{u,i}$ of that item.
Since these components are precomputed during the previous persona construction step, assembling user profiles is efficient and requires no additional reasoning.

For each observed interaction $(u, i)$, we employ LLM-as-a-judge approach to select the persona from the target item persona set 
$\Pi_i$ that best explains why the user engaged with the item.
Given the user profile $P_u$, the target item title $t_i$, and its persona set $\Pi_i$, the LLM $\mathcal{M}$, guided by an alignment instruction $\mathcal{I}_{\mathrm{align}}$, evaluates their semantic alignment and outputs the most relevant persona $\pi_{(u,i)}^+ $ along with a concise justification $j_{(u,i)}^+$:

\begin{equation}
\mathcal{M}:(P_u, t_i, \Pi_i, \mathcal{I}_{\mathrm{align}}) \to \big(j_{(u,i)}^+, \pi_{(u,i)}^+\big)
\label{eq:judge}
\end{equation}

This process links each user's historical preferences with the most representative persona of the target item,
yielding the alignment dataset $\mathcal{D}_{\text{align}} = \{ (u,i,\pi_{(u,i)}^+) \}$ that converts implicit interactions into explicit, interpretable supervision for encoder training.

\subsubsection{Encoder Training}
\label{subsubsec:encoder_training}
This step trains a lightweight encoder to embed user profiles and personas into a shared vector space, enabling efficient similarity-based scoring.
Using the alignment dataset $\mathcal{D}_{\text{align}}$, we train an encoder $E_{\theta}$ via contrastive learning.

\smallsection{User Profile \& Persona Embedding}  
We encode both user profiles and personas using the same encoder $E_{\theta}$. 
For a user $u$, we encode the $l$-th most recent interaction as:
\begin{equation}
    e_{u,l} = E_{\theta}(\text{concat}(s_{i_l}, a_{u,i_l})),
\end{equation}
where $l \in \{1, \ldots, L\}$ and $i_l$ denotes the item at position $l$. To obtain a user embedding, we aggregate interaction-level embeddings with a temporal decay factor to account for recency effects:
\begin{equation}
    e_u = \frac{\sum_{l=1}^{L} \gamma^{l-1} \cdot e_{u,l}}{\sum_{l=1}^{L} \gamma^{l-1}},
\end{equation}
where $\gamma \in (0,1]$ controls the influence of older interactions.
Similarly, each persona $\pi \in \Pi_i$ is embedded as
\begin{equation}
    e_{\pi} = E_{\theta}(\pi),
\end{equation}

\smallsection{Contrastive Learning}  
We optimize the encoder on the alignment dataset $\mathcal{D}_{\text{align}}$ using the InfoNCE objective~\cite{oord2018representation}:
\begin{equation}
    \mathcal{L} = -\log 
    \frac{\exp(\mathrm{sim}(e_{u}, e_{\pi^+}) / \tau)}
    {\sum_{j=1}^{B} \exp(\mathrm{sim}(e_{u}, e_{\pi_j^-}) / \tau)},
\label{eq:infonce}
\end{equation}
where $B$ is the batch size, $\tau$ is the temperature, 
with in-batch negatives.
This contrastive objective trains the encoder to approximate the LLM-derived alignment supervision by mapping aligned user–persona pairs closer in the embedding space than mismatched pairs.
{After training, all item personas are encoded once to build a persona-profiled item index,}
enabling efficient similarity-based scoring during inference without additional LLM reasoning.

\subsection{Online Inference}
\label{subsec:inference}

The online stage 
of \proposed\ enables efficient real-time recommendation by combining precomputed persona embeddings $e_{\pi}$ with user profiles $P_u$ composed from the user’s latest interactions.
Since all personas are indexed offline, online inference avoids LLM invocation and requires only lightweight user encoding and similarity scoring against cached embeddings.

\smallsection{User Encoding}  
At the online 
stage, the user embedding $e_u$ is computed by aggregating interaction-level embeddings through the same temporally weighted scheme as in training.
Our method operates with as few as one interaction, though richer histories enable more precise persona alignment. When a new interaction occurs with an associated review, the aspect is extracted and cached asynchronously to avoid latency overhead.

\smallsection{Candidate Item Scoring and Reranking}  
For each candidate item $i \in \mathcal{C}_u$, the relevance score is computed as the maximum similarity between the user embedding and the item’s personas:
\begin{equation}
    \text{score}(u, i) = \max_{\pi \in \Pi_i} \mathrm{sim}(e_u, e_\pi).
\end{equation}
Items are reranked by this score, and the persona achieving the maximum similarity serves as a human-interpretable explanation by providing its \textit{description} and \textit{preference rationale.}


\input{TBL/043_dataset_status}
\vspace{-0.2cm} 
\begin{figure}[t]
    \centering
    \includegraphics[width=\linewidth]{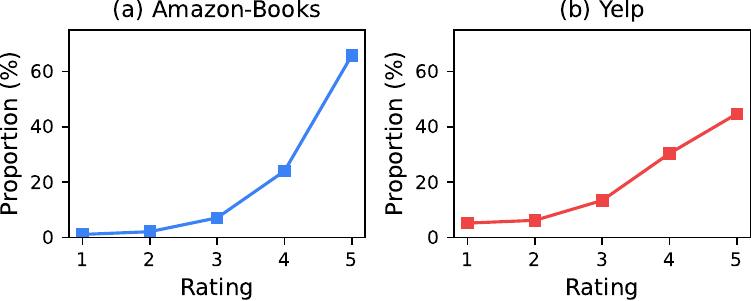}
    \caption{Rating distribution of the two dataset}
    \label{fig:Data}
\end{figure}

%% file: TBL/033_Persona_example.tex
\begin{table}[t]
\centering
\renewcommand{\arraystretch}{0.9}
\rmfamily   
\caption{Example of persona construction for item {``The Shadow in the Glass''}---a dark retelling of Cinderella. }
\vspace{-0.2cm}
\label{tab:persona-example-dark}
\begin{tabular}{p{0.95\linewidth}}
\toprule

\textbf{Review Aspects (user \textit{AEPTNCI3X5})}\\
\midrule
\textbf{Category Preference:} {Dark fantasy, gothic retellings}\\
\textbf{Purchase Purpose:} {Interest in morally complex reinterpretations of classic fairy tales}\\
\textbf{Quality Criteria:} {Tension, surprising twists, and morally ambiguous characters}\\
\textbf{Usage Context:} {Immersive reading during leisure hours}\\
\midrule
\textbf{Constructed Persona}\\

\midrule
\textbf{Name:} {The Dark Storyline Seeker}\\
\textbf{Description:} 
{This reader is attracted to darker storylines filled with suspense and moral complexity. 
They enjoy unpredictable twists and flawed characters whose motives remain uncertain. 
Rather than seeking comfort, they appreciate stories that challenge expectations and evoke unease.}\\
\textbf{Preference Rationale:} 
{This persona appreciates this item because they enjoy a darker storyline with surprises 
and morally questionable characters, as frequently mentioned in reviews describing the story’s tension and sense of danger.}\\
\bottomrule
\end{tabular}
\end{table}

%% file: TBL/043_dataset_status.tex
\begin{table}[t]
\centering
\small  
\caption{Statistics of the datasets.}
\vspace{-0.2cm}
\label{tab:dataset-summary}
\resizebox{0.99\linewidth}{!}{
\begin{tabular}{lcccc}
\toprule
\textbf{Dataset} & \textbf{\#Interactions} & \textbf{\#Users} & \textbf{\#Items}  & \textbf{Sparsity} \\
\midrule
\amazon & 309,287 & 26,173 & 25,130 & 99.9530\%
 \\
\yelp & 103,774 & 7,968 & 2,942 & 99.5573\%
 \\
\bottomrule
\end{tabular}
}
\end{table}

%% file: TEX/040experiments_WWW.tex
\input{TBL/040_RQ1-1_main_results}
In this section, we design and conduct experiments to answer the following research questions.
\begin{itemize}[leftmargin=1.2em, itemsep=2pt, topsep=2pt]
  \item \textbf{RQ1 (Reranking Performance):} 
  How well does \proposed rerank candidates produced by first-stage recommenders, 
  and how robust is it across sparsity scenarios?

  \item \textbf{RQ2 (Efficiency \& Scalability):} 
  How do inference latency and memory footprint 
  compare to LLM-based rerankers, and how do they scale with request volume?

  \item \textbf{RQ3 (Recommendation
Explainability):}  
  How effectively do persona-based rationales convey clear, consistent, and engaging explanations of recommendations?

  \item \textbf{RQ4 (Effect of Multi-Persona Modeling):} 
  How well does modeling multiple personas per item capture diverse and fine-grained user motivations for engagement?
\end{itemize}

\subsection{Experimental Settings}

\subsubsection{Datasets} 
To evaluate the effectiveness and robustness of \proposed\ across domains, we experiment with two real-world datasets: \textbf{\amazon}~\cite{hou2024bridging} 
and \textbf{\yelp}.\footnote{https://www.yelp.com/dataset/}

\begin{itemize}[leftmargin=1.2em, itemsep=2pt, topsep=2pt]
    \item{\textbf{\amazon}}: 
    We use the Books category from the Amazon Reviews 2023 dataset~\cite{hou2024bridging}. 
    The dataset provides user ratings on a 1-5 scale, along with textual reviews and metadata.

    \item{\textbf{\yelp}}: We focus on the restaurant and food categories in the Philadelphia region. 
    Similar to \amazon, it provides user ratings on a 1-5 scale, textual reviews, and metadata.
\end{itemize}
Following established practices in LLM-based recommendation~\cite{bao2023tallrec,jackermeier2024dual,kim2025driven,yue2023llamarec}, we subsample approximately 100K-300K interactions to manage the computational cost of LLM-based methods,
and apply both metadata and 5-core filtering to ensure data quality.
The detailed statistics are summarized in Table~\ref{tab:dataset-summary} and Figure~\ref{fig:Data}.

\subsubsection{Baselines}
For the top-$K$ item recommendation task, we adopt a multi-stage ranking pipeline, where an initial candidate set is generated by a first-stage retriever and subsequently refined by a second-stage reranker.
Within this framework, our main baselines are \textbf{LLM-based rerankers}, which directly apply LLMs during inference to reorder candidates. 
We compare \proposed against two widely adopted LLM-based rerankers, \zsllm~\cite{sun2023chatgpt} and \tallrec~\cite{bao2023tallrec}, as well as the recent state-of-the-art method \expert~\cite{kim2025driven}. 
For the \textbf{first-stage candidate retrievers}, we use CF models including \bprmf~\cite{rendle2012bpr} and \lightgcn~\cite{he2020lightgcn}. Additionally, for explainability comparison, we include two \textbf{review-based rationale generators}: \xrec~\cite{ma2024xrec} and \expert~\cite{kim2025driven}. 

\subsubsection{Evaluation Protocol}

For our main evaluation, we focus on the top-$K$ reranking task and report HR@$K$, MRR@$K$, and NDCG@$K$.
for $K=\{5, 10, 20\}$ as the evaluation metric,
following the established practices in previous work \cite{yue2023llamarec,kim2025driven}. 
We adopt leave-one-out (LOO) splitting for train/validation/test~\cite{sun2019bert4rec,yue2023llamarec,zhou2020s3}.   
\input{TBL/041_RQ1-2_Scenario_Reranking}
\subsubsection{Implementation Details}
Following the existing convention of constructing user profile~\cite{bao2023tallrec}, we set history length $l=10$, and gamma is set through validation.
For encoder training, we fine-tune the BGE-M3~\cite{chen2024bgem3embeddingmultilingualmultifunctionality} model using LoRA~\cite{hu2022lora} with rank $r=16$, alpha $\alpha=32$, and dropout $0.05$. 
We use a global batch size of 72 and learning rate of $1\times10^{-5}$ across all experiments.

\subsection{RQ1: Reranking Performance}

We first examine whether \proposed can effectively enhance top-$K$ recommendation accuracy when used to rerank candidates produced by CF models, to verify that shifting reasoning offline does not compromise ranking quality. 
\input{TBL/046_RQ2_summary_only}
\subsubsection{Overall Performance.}  

Table~\ref{tab:main-rerank} shows \proposed consistently outperforms 
CF generators and LLM-based rerankers across datasets. Notably, LLM-based rerankers struggle under real-world data conditions, with some falling behind CF generators due to dataset-specific limitations. 
In \amazon, extreme rating skew toward 5-stars {(Figure~\ref{fig:Data})} undermines \expert;
when most items share similar ratings, rating-based reasoning loses discriminative power. 
On the other hand, \zsllm's generic prompting and \tallrec's instruction-tuned approach fail to capture nuanced collaborative patterns. 
In \yelp, rating skew is less extreme than \amazon but still exhibits significant imbalance (Figure~\ref{fig:Data}). When this unbalanced distribution is combined with sparse metadata (e.g., ``WiFi available,'' ``child-friendly''), it provides insufficient differentiation for reasoning-based methods.

\proposed overcomes these limitations by extracting personas primarily
from review contents rather than focus on metadata or ratings. While LLM rerankers are affected by noisy raw metadata for reasoning, we refine this noisy metadata into structured summaries. This refinement process provides more objective and dense information for constructing personas, even when metadata is sparse. 
Besides, even though ratings are skewed, personas can capture diverse aspects from review semantics. 
This allows our model to avoid the loss of discriminative power that hinders rating-based reasoning approaches.
Beyond this textual refinement, our framework learns to align user interaction histories with these personas through contrastive learning, enabling the model to capture implicit behavioral preferences. 
This alignment process allows the encoder to recognize subtle preference patterns embedded in users' actual engagement histories. 
This synergy between semantic understanding and behavioral alignment ensures robust and reliable performance even under challenging data conditions. Even without fine-tuning, the vanilla variant with pre-trained BGE-M3 outperforms LLM baselines, validating the effectiveness of our persona-based textual refinement. Fine-tuning adds further gains by learning from interaction patterns through user-persona alignment training.

\subsubsection{Cold-/Warm-start Scenario Analysis}
We partition test interactions by review count into user/item segments to test robustness under sparsity (warm/head: top 20\%, cold/tail: bottom 20\%).
As shown in Table~\ref{tab:cold-warm-analysis}, \proposed exhibits gains across all regimes.

\noindent \textbf{Tail items deliver the suitable gains.}
These items remain sparsely connected within the dataset, causing CF generators to produce unreliable rankings from weak collaborative patterns. In contrast, \proposed 
infers structured personas from the available reviews—even when reviews are few, they provide richer semantic signals than interaction counts alone. This review-grounded evidence enables robust reranking where CF struggles most.

\noindent \textbf{Warm users show strong relative improvements.}
Despite starting from low baseline performance, warm users achieve substantial relative gains. 
This suggests that persona-based matching provides significant value 
when users have extensive histories, as richer context enables alignment 
with more specialized personas that capture preference nuances invisible 
to collaborative filtering. 

\noindent \textbf{Head items show modest gains.}
Two factors may contribute to this ceiling. First, abundant interactions enable CF to learn robust patterns, reducing the marginal value of semantic augmentation. Second, popular items often exhibit broader preference diversity
than what our current persona generation framework can capture.
Although \proposed limits each item to a maximum of seven personas for computational efficiency,
highly reviewed items might benefit from a more fine-grained representation.

\noindent \textbf{Cold users show moderate improvements.}
Cold users exhibit smaller relative improvements as sparse histories limit coherent preference profiling—especially when consuming popular items with generic taste signals. 
In contrast, warm users' extensive histories reveal consistent patterns (e.g., recurring themes, quality criteria) enabling precise alignment with specialized personas. 
This confirms that persona semantics provide value across sparsity levels, but optimal matching requires sufficient interaction data.



\subsubsection{Robustness to Missing Reviews}
A common limitation of review-based methods is their inability to handle items lacking reviews, which frequently occurs for newly listed items in real-world platforms.
To evaluate robustness when items have metadata but no reviews yet, we simulate this by using the same trained encoder but indexing items with their summary $e_i = E_\theta(s_i)$ instead of personas (Summary Only).
As shown in Table~\ref{tab:ablation-summary}, this variant still outperforms the CF baseline across all settings, demonstrating that our framework can effectively serve cold-start items before reviews accumulate, as the encoder has already learned summary semantics through user profiles during training. When reviews become available, the Full model yields additional gains, confirming the complementary value of review-grounded persona reasoning.

Overall, \proposed demonstrates robust improvements across all data —including complete cold-start scenarios where no reviews are available—confirming our method effectively complements CF-based candidate generation under diverse conditions.

\input{TBL/044_RQ2-1_Latency}
\begin{figure}[t]
\vspace{-0.2cm}
    \centering
    \includegraphics[width=0.98\linewidth]{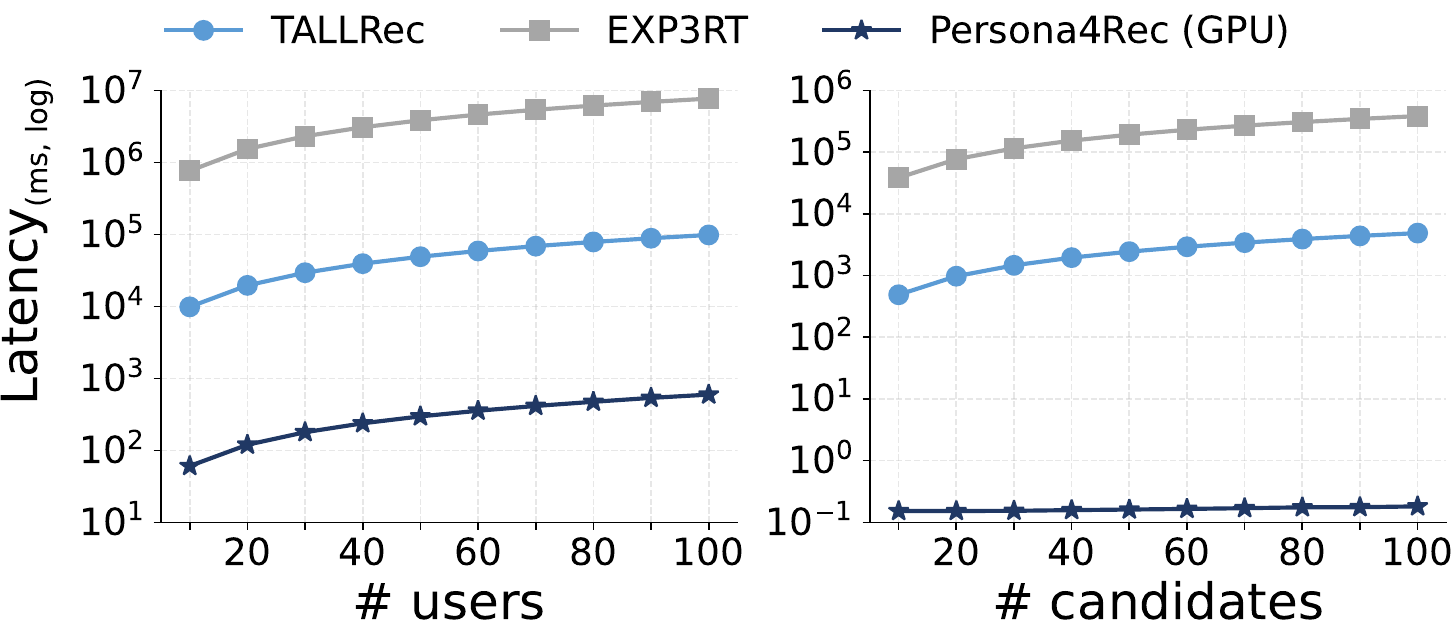}
    \vspace{-0.1cm}
    \caption{Comparison of inference scalability with respect to the number of user samples and candidate items.}
    \label{fig:RQ2-Scalability}
\end{figure}

\subsection{RQ2: Efficiency \& Scalability}
We evaluate inference efficiency on a single A6000 GPU, 
measuring latency and memory usage against LLM-based 
baselines.

\subsubsection{Single-Query Performance.}
In Table~\ref{tab:efficiency_comparison}, 
\proposed achieves 0.75ms latency (GPU) or 0.52ms (CPU), representing \textbf{1,300$\times$ speedup over \tallrec and 104,000$\times$ over \expert}. Memory footprint is similarly reduced: 0.1GB versus 12--15GB for baselines. This efficiency stems from replacing LLM inference with dot-product operations over pre-computed persona embeddings.

\subsubsection{Scalability Evaluation.}
Beyond single-query speed, we assess how \proposed scales 
under realistic deployment conditions—a critical requirement 
for production recommender systems serving millions of users. 
Figure~\ref{fig:RQ2-Scalability} evaluates two dimensions: 
batch size (concurrent user requests) and item candidate set size.

\noindent \textbf{User scaling.} 
In Figure~\ref{fig:RQ2-Scalability} Left,
as the number of concurrent users increases from 1 to 100, 
LLM-based methods exhibit superlinear growth due to 
sequential processing constraints. 
\tallrec's latency grows from $10^4$ms to $10^5$ms 
($\sim$100 seconds), while \expert deteriorates from 
$10^6$ms to over $10^7$ms (exceeding 2 hours). 
In contrast, \proposed's latency increased from $10^2$ms to only $10^3$ms. 

\noindent \textbf{Candidate scaling.}
In Figure~\ref{fig:RQ2-Scalability} Right,
when the candidate set size grows from 20 to 100 items, 
\proposed maintains near-constant latency (flat at $\sim$1ms) 
via efficient vector operations. 
\tallrec scales moderately due to per-item reasoning costs, 
while \expert remains prohibitively slow regardless of 
candidate count. 
This insensitivity to candidate size enables reranking without latency penalties.

Overall, these results confirm that \proposed not only achieves superior efficiency but also exhibits fundamentally different scaling characteristics. The architectural advantage of offline persona construction is amplified under high load, making real-time recommendation at scale practical with lower computational cost.
\input{TBL/062_humaneval-RQ3}

\subsection{RQ3: Recommendation Explainability}
\label{subsec:RQ3}  
We evaluate whether persona-based rationales effectively explain recommendations through pairwise comparisons against \expert (online reasoning) and \xrec (single-review summary).
We evaluate five user-facing criteria that capture the explanatory effectiveness: 
\textit{Intuitiveness}, 
\textit{Persuasiveness}, 
\textit{Relatability}, \textit{Consistency}, and \textit{Specificity}.
These collectively measure the explainability of recommendation rationales from the target user's perspective.
(Refer to Table~\ref{tab:rq3-criteria} for more details.)
Following prior work~\cite{kim2025driven, kim2024pearl}, we sample 100 examples per dataset. To ensure objectivity, we employed impartial annotators via Amazon Mechanical Turk (AMT), assigning three independent annotators per example.

\begin{figure}[t]
    \centering
    \includegraphics[width=\linewidth]{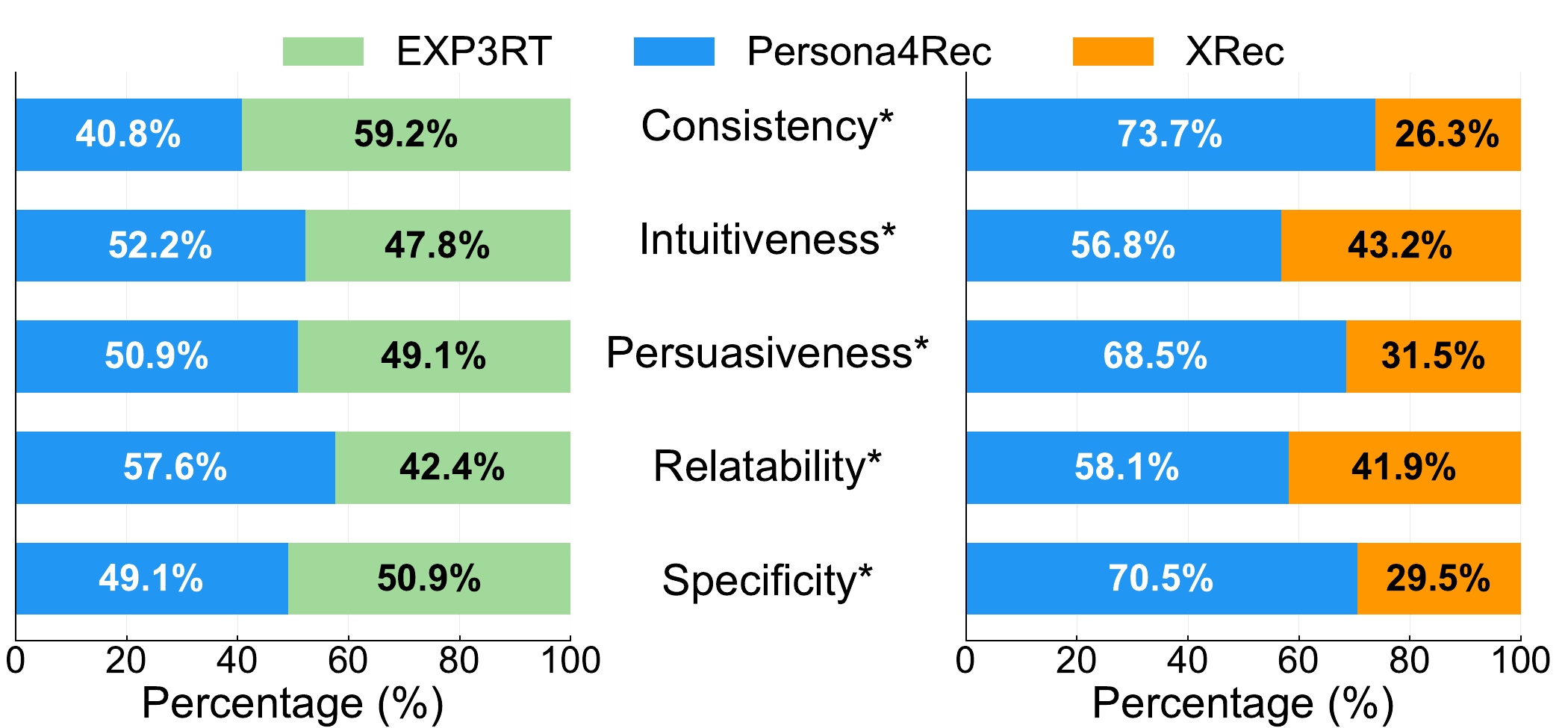}
    \vspace{-15pt}
    \caption{Human evaluation of Recommendation Explainability across compared methods. ($*$: p-value < 0.05)}
    \label{fig:Persona raionlae}
    \vspace{-5pt}
\end{figure}
\input{TBL/062_humaneval-RQ4}

In Figure~\ref{fig:Persona raionlae}, \proposed achieves near-parity with \expert while dominating \xrec.
Against \expert, we trade specificity for efficiency structure. \expert's real-time reasoning enables highly personalized rationales tailored to individual contexts, whereas our pre-constructed personas represent aggregated patterns. 
However, we match \expert in intuitiveness and persuasiveness through structured formatting consistent persona templates, which enhance comprehensibility and review-grounded evidence that provides tangible social proof. 
Critically, this balanced trade-off delivers inference speedup, making real-time deployment practical.
Compared to \xrec, our approach dominates across all dimensions, including \textit{specificity}. \proposed extracts 2--7 review-grounded personas per item, each capturing a distinct user motivation, while \xrec represents each item with a single item profile that aggregates all review signals, limiting its ability to distinguish between different user motivations.
Overall, \proposed delivers effective explainability  through offline pre-defined personas.

\subsection{RQ4: Effect of Multi-Persona Modeling}
We now examine how the number of personas per item ($M$) influences both the quality of the constructed personas and their impact on recommendation performance. 
Specifically, we analyze whether increasing $M$ enhances the faithfulness and clarity of personas and leads to greater improvements in recommendation accuracy.

\begin{figure}[t]
    \centering
    \includegraphics[width=\linewidth]{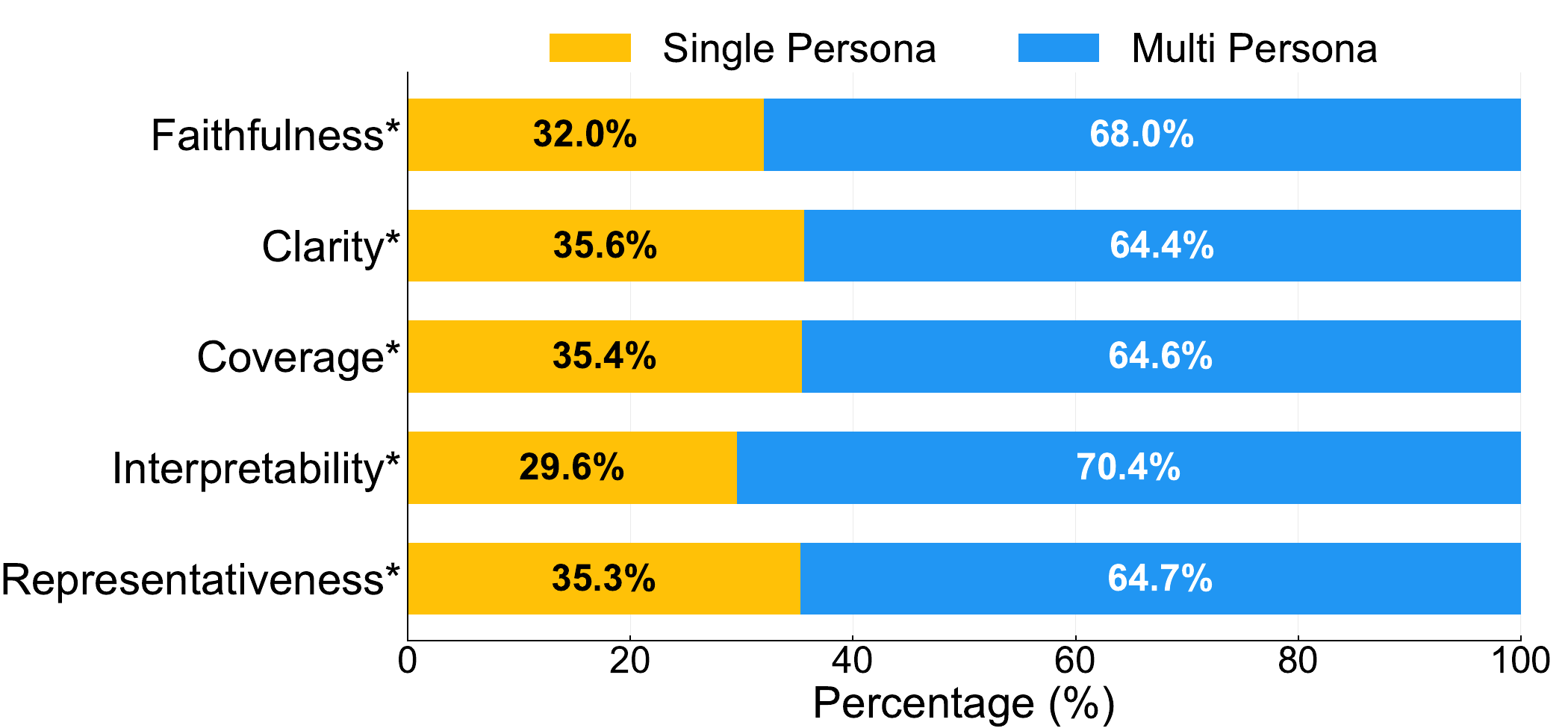}
    \vspace{-15pt}
    \caption{Human evaluation on the persona construction quality in single- and multi-persona. ($*$: p-value < 0.05)}
    \label{fig:Persona Granularity}
    \vspace{-0.8cm}
\end{figure}

\subsubsection{Effect of the Number of Personas on Quality}
We compare \textit{Single-Persona} (one aggregated persona) with \textit{Multi-Persona} (our default, $K\in[2,7]$) using human evaluation on five criteria. 
We adopt these five criteria that cover the key aspects of persona quality in the construction stage: 
\textit{Faithfulness}, \textit{Clarity}, \textit{Coverage}, 
\textit{Interpretability}, and \textit{Representativeness}. 
(Refer to Table~\ref{tab:rq4-criteria} for more details.)

\input{TBL/045_RQ4_Table}

As shown in Figure~\ref{fig:Persona Granularity}, Multi-persona generation achieved substantially higher scores across all five evaluation criteria, with improvements of over 30 percentage points in faithfulness and interpretability. These results confirm that aggregating all signals into a single persona leads to oversimplification and loss of nuance, whereas modeling items with multiple distinct personas provides more consistent and specific rationales that better reflect heterogeneous user motivations.
\vspace{-5pt}

\subsubsection{Effect of the Number of Personas on  Performance}
Moreover, to evaluate how the number of personas per item influences recommendation performance, 
we design an ablation study on the \amazon dataset using BPR-MF as the candidate generator. 
For each item, we vary the number of generated personas (1, 3, 5, and 7) and use them for reranking evaluation, 
reporting HR and NDCG at cutoff levels. 
This experiment is conducted on the 13,302 users whose candidate item persona pool contains exactly seven personas.
By restricting evaluation to this setting, we can directly compare the impact of the number of personas (1--7) while keeping the candidate pool consistent.
As shown in Table~\ref{tab:persona-ablation}, increasing the number of personas consistently improves ranking performance, 
with the best results observed at 7 personas. 
Even a moderate number of personas
(e.g., 5 personas) yields steady gains, 
demonstrating that representing items with multiple distinct personas is more effective than compressing all review signals into a single aggregated profile.
Overall, these findings suggest that \textbf{multi-persona generation provides a finer-grained profiling of user motivations}, 
capturing both majority and minority perspectives.  
This richer representation not only improves ranking metrics but also enhances the explainability of recommendations by reflecting diverse user preferences across heterogeneous groups.

%% file: TBL/040_RQ1-1_main_results.tex
\begin{table*}[t]
\centering
\footnotesize
\setlength{\tabcolsep}{3pt}
\renewcommand{\arraystretch}{0.98}
\caption{Top-$K$ recommendation performance of \proposed and other LLM-based reranking methods on both datasets. For each user, candidate items are initially retrieved by two CF models: \bprmf and \lightgcn. ( * : p < 0.05 )}
\vspace{-2mm}
\label{tab:main-rerank}
\resizebox{0.99\linewidth}{!}{
\begin{tabular}{cc l ccc ccc ccc}
\toprule
\multirow{2.5}{*}{\textbf{Dataset}} & 
\multirow{2.5}{*}{\textbf{Generator}} & 
\multirow{2.5}{*}{\textbf{Reranker}} &
\multicolumn{3}{c}{\textbf{@5}} &
\multicolumn{3}{c}{\textbf{@10}} &
\multicolumn{3}{c}{\textbf{@20}} \\
\cmidrule(lr){4-6}\cmidrule(lr){7-9}\cmidrule(lr){10-12}
& & & \textbf{HR} & \textbf{MRR} & \textbf{NDCG} & 
     \textbf{HR} & \textbf{MRR} & \textbf{NDCG} & 
     \textbf{HR} & \textbf{MRR} & \textbf{NDCG} \\
\midrule
\multirow{12}{*}{\rotatebox[origin=c]{90}{\textbf{\amazon}}}
& \multirow{6}{*}{\textbf{\bprmf}}
& -                     & 0.0639 & 0.0340 & 0.0414 & 0.1000 & 0.0388 & 0.0531 & & 0.0419 & 0.0643 \\
& & \zsllm              & 0.0664 & 0.0357 & 0.0433 & 0.1025 & 0.0405 & 0.0550 & $\vert$ & 0.0434 & 0.0655 \\
& & \tallrec            & 0.0346 & 0.0166 & 0.0210 & 0.0667 & 0.0207 & 0.0313 & \multirow{2}{*}{0.1446} & 0.0258 & 0.0505 \\
& & \expert             & 0.0498 & 0.0238 & 0.0302 & 0.0865 & 0.0286 & 0.0420 & & 0.0326 & 0.0565 \\
& & \proposed (vanilla) 
& \ul{0.0668} & \textbf{0.0389} & \ul{0.0457} 
& \ul{0.1007} & \ul{0.0433} & \ul{0.0567} 
& {$\vert$} 
& \textbf{0.0464} & \ul{0.0678} \\
& & \proposed (fine-tuned)             & \textbf{0.0707}$^{*}$ & \ul{0.0385}&\textbf{0.0464} & \textbf{0.1081}$^{*}$ & \textbf{0.0434} & \textbf{0.0585}$^{*}$ & & \ul{0.0460} & \textbf{0.0679} \\
\cmidrule{2-12}
& \multirow{6}{*}{\textbf{LightGCN}}
& -                     & 0.0671 & 0.0353 & 0.0432 & 0.1071 & 0.0406 & 0.0561 & & 0.0440 & 0.0681 \\
& & \zsllm              & 0.0673 & 0.0364 & 0.0440 & 0.1071 & 0.0417 & 0.0568 & $\vert$ & 0.0450 & 0.0690 \\
& & \tallrec            & 0.0365 & 0.0166 & 0.0215 & 0.0716 & 0.0211 & 0.0327 & \multirow{2}{*}{0.1555} & 0.0265 & 0.0534 \\
& & \expert             & 0.0455 & 0.0213 & 0.0273 & 0.0828 & 0.0262 & 0.0392 & & 0.0311 & 0.0574 \\
& & \proposed (vanilla) & \ul{0.0696} & \ul{0.0401} & \ul{0.0474} & \ul{0.1055} & \ul{0.0449} & \ul{0.0590}
& {$\vert$} & \ul{0.0483} & \ul{0.0716} \\
& &\proposed (fine-tuned)             & \textbf{0.0757}$^{*}$ & \textbf{0.0410} & \textbf{0.0495}$^{*}$ & \textbf{0.1151}$^{*}$ & \textbf{0.0462}$^{*}$ & \textbf{0.0622}$^{*}$ & &   \textbf{0.0491}$^{*}$ & \textbf{0.0726}$^{*}$ \\
\midrule
\multirow{12}{*}{\rotatebox[origin=c]{90}{\textbf{\yelp}}}
& \multirow{6}{*}{\textbf{\bprmf}}
& -   & 0.0334 & 0.0157 & 0.0200 & \ul{0.0557} & 0.0186 & 0.0272 &  & 0.0213 & 0.0370 \\
& & \zsllm              & 0.0285 & 0.0132 & 0.0169 & 0.0499 & 0.0160 & 0.0238 & $\vert$ & 0.0191 & 0.0351 \\
& & \tallrec            & 0.0278 & 0.0125 & 0.0162 & 0.0537 & 0.0159 & 0.0245 & \multirow{2}{*}{0.0949} & 0.0187 & 0.0349 \\
& & \expert             & 0.0265 & 0.0110 & 0.0148 & 0.0516 & 0.0141 & 0.0227 & & 0.0171 & 0.0336 \\
& & \proposed (vanilla) & \ul{0.0349} & \ul{0.0175} & \ul{0.0217} & 0.0555 & \ul{0.0201} & \ul{0.0283} & {$\vert$} & \ul{0.0228} & \ul{0.0382} \\
& & \proposed (fine-tuned)    & \textbf{0.0351}$^{*}$ & \textbf{0.0181}$^{*}$ & \textbf{0.0223}$^{*}$ & \textbf{0.0599}$^{*}$ & \textbf{0.0214}$^{*}$ & \textbf{0.0303}$^{*}$ & & \textbf{0.0238}$^{*}$ & \textbf{0.0391}$^{*}$ \\
\cmidrule{2-12}
& \multirow{6}{*}{\textbf{\lightgcn}}
& -                     & 0.0388 & 0.0195 & 0.0242 
& \ul{0.0682} & 0.0233 & \ul{0.0336} 
& & 0.0258 & 0.0430 \\
& & \zsllm              & \textbf{0.0398} &\textbf{0.0201} & \textbf{0.0249} & 0.0664 & \ul{0.0236} & 0.0334 & $\vert$ & \ul{0.0262} & \ul{0.0432} \\
& & \tallrec            & 0.0325 & 0.0147 & 0.0191 & 0.0616 & 0.0185 & 0.0284 & \multirow{2}{*}{0.0106} & 0.0215 & 0.0394 \\
& & \expert             & 0.0292 & 0.0125 & 0.0166 & 0.0561 & 0.0160 & 0.0252 & & 0.0193 & 0.0376 \\
& & \proposed (vanilla) 
& 0.0351 & 0.0180 & 0.0222 & 0.0595 & 0.0211 & 0.0300 & {$\vert$} & 0.0242 & 0.0416 \\
& & \proposed (fine-tuned) 
& \ul{0.0395} & \ul{0.0199} & \ul{0.0247} 
& \textbf{0.0694} & \textbf{0.0238} & \textbf{0.0343} 
& & \textbf{0.0264} & \textbf{0.0436} \\
\bottomrule
\end{tabular}
}
\end{table*}

%% file: TBL/041_RQ1-2_Scenario_Reranking.tex
\begin{table*}[t]
\centering
\scriptsize
\setlength{\tabcolsep}{7pt} 
\renewcommand{\arraystretch}{1.01} 
\caption{Reranking performance across cold-start and warm-start scenarios on \amazon dataset, with relative improvements (\%) over CF baseline, denoted by ``-''. ($*$: p-value < 0.05)} 
\vspace{-0.2cm}
\label{tab:cold-warm-analysis}
\resizebox{\linewidth}{!}{
\begin{tabular}{clcccccccc}
\toprule
\multirow{3}{*}{\textbf{Generator}} & 
\multirow{3}{*}{\textbf{Reranker}} & \multicolumn{2}{c}{\textbf{Warm Users@10}} & \multicolumn{2}{c}{\textbf{Cold Users@10}} & 
     \multicolumn{2}{c}{\textbf{Head Items@10}} & \multicolumn{2}{c}{\textbf{Tail Items@10}} \\
\cmidrule(lr){3-4}\cmidrule(lr){5-6}\cmidrule(lr){7-8}\cmidrule(lr){9-10}
& & \textbf{HR} & \textbf{NDCG} & \textbf{HR} & \textbf{NDCG} & 
     \textbf{HR} & \textbf{NDCG} & \textbf{HR} & \textbf{NDCG} \\
\midrule
\multirow{7}{*}{\textbf{\bprmf}}
& - & 0.0469 & 0.0236 & 0.1380 & 0.0757 & 0.1662 & 0.0868 & 0.0310 & 0.0185 \\
& \zsllm      
& \ul{0.0490} & \ul{0.0251} & \ul{0.1409} & \ul{0.0778} & \ul{0.1696} & \ul{0.0897} & \ul{0.0323} & \ul{0.0191} \\
& \tallrec     & 0.0451 & 0.0216 & 0.0749 & 0.0333 & 0.1120 & 0.0530 & 0.0160 & 0.0076 \\
& \expert      & 0.0486 & 0.0236 & 0.1109 & 0.0528 & 0.1457 & 0.0683 & 0.0268 & 0.0160 \\
& \textbf{\proposed (ours)} & \textbf{0.0599}$^{*}$ & \textbf{0.0320}$^{*}$ & \textbf{0.1432} & \textbf{0.0789} & \textbf{0.1707} & \textbf{0.0904} & \textbf{0.0360}$^{*}$ & \textbf{0.0196} \\
\cmidrule{2-10}
& \textit{improv. over BPR-MF} & \textit{+27.7\%} & \textit{+35.6\%} & \textit{+3.8\%} & \textit{+4.2\%} & \textit{+2.7\%} & \textit{+4.1\%} & \textit{+16.1\%} & \textit{+5.9\%} \\
\midrule
\multirow{7}{*}{\textbf{\lightgcn}}
& - & \ul{0.0552} & 0.0267 & \ul{0.1479} & 0.0782 & \ul{0.1671} & \ul{0.0858} & 0.0383 & 0.0225 \\
& \zsllm      
& 0.0536 & \ul{0.0273} & 0.1474 & \ul{0.0803} &  0.1658& 0.0858 & \ul{0.0385} & \ul{0.0235} \\
& \tallrec     & 0.0462 & 0.0213 & 0.0841 & 0.0363 & 0.1124 & 0.0519 & 0.0235 & 0.0106 \\
& \expert      & 0.0445 & 0.0211 & 0.1049 & 0.0495 & 0.1295 & 0.0591 & 0.0330 & 0.0170 \\
& \textbf{\proposed (ours)} & \textbf{0.0621}$^{*}$ & \textbf{0.0328}$^{*}$ & \textbf{0.1545}$^{*}$ & \textbf{0.0850}$^{*}$ & \textbf{0.1697}$^{*}$ & \textbf{0.0905}$^{*}$ & \textbf{0.0455}$^{*}$ & \textbf{0.0233} \\
\cmidrule{2-10}
& \textit{improv. over LightGCN} & \textit{+12.5\%} & \textit{+22.8\%} & \textit{+4.5\%} & \textit{+8.7\%} & \textit{+1.6\%} & \textit{+5.5\%} & \textit{+18.8\%} & \textit{+3.6\%} \\
\bottomrule
\end{tabular}

}
\end{table*}

%% file: TBL/046_RQ2_summary_only.tex
\begin{table}[t]
\centering
\small
\caption{Simulating items without reviews on \amazon. Results at @10 with relative improvement over CF baseline.}
\vspace{-0.2cm}
\label{tab:ablation-summary}
\begin{tabular*}{\linewidth}{@{\extracolsep{\fill}}clll}
\toprule
\textbf{\hspace{2mm}Generator} & \textbf{Reranker} & \multicolumn{1}{c}{\textbf{HR}} & \multicolumn{1}{c}{\textbf{NDCG}} \\
\midrule
\multirow{3}{*}{\hspace{2mm}\parbox{1.5cm}{\centering\textbf{\bprmf}}}
& - & 0.1000 & 0.0531 \\
& Summary Only & \ul{0.1043} \textit{(+4.3\%)} & \ul{0.0560} \textit{(+5.5\%)} \\
& Full (Ours) & \textbf{0.1081} \textit{(+8.1\%)} & \textbf{0.0585} \textit{(+10.2\%)} \\
\midrule
\multirow{3}{*}{\hspace{2mm}\parbox{1.5cm}{\centering\textbf{\lightgcn}}}
& - & 0.1071 & 0.0561 \\
& Summary Only & \ul{0.1112} \textit{(+3.8\%)} & \ul{0.0591} \textit{(+5.3\%)} \\
& Full (Ours) & \textbf{0.1151} \textit{(+7.5\%)} & \textbf{0.0622} \textit{(+10.9\%)} \\
\bottomrule
\end{tabular*}
\vspace{-0.2cm}
\end{table}

%% file: TBL/044_RQ2-1_Latency.tex
\begin{table}[t]
\caption{Comparison of inference efficiency between LLM-based rerankers and \proposed. Latency for \zsllm is measured as end-to-end API latency.}
\vspace{-0.2cm}
\label{tab:efficiency_comparison}
\centering
\resizebox{0.99\linewidth}{!}{
\begin{tabular}{ccc}
\toprule
\textbf{Method} 
& {\textbf{Inference Latency (ms)}} 
& {\textbf{Memory Footprint (GB)}} \\
\midrule
\zsllm            & 5,537 $\pm$ 3,094.13 & - \\
\tallrec          & 979.17 $\pm$ 13.29 & 12.57 \\
\expert           & 77,974.08 $\pm$ 1,229.99 & 14.96 \\
\midrule
\multirow{2}{*}{\textbf{\proposed}} & \textbf{0.52 $\pm$ 0.01 (CPU)} & - \\
 & \textbf{0.75 $\pm$ 0.02 (GPU)} & \textbf{0.1} \\
\bottomrule
\end{tabular}
}
\end{table}

%% file: TBL/062_Humaneval-RQ3.tex
\begin{table}[t]
\centering
\small
\setlength{\tabcolsep}{4pt}
\renewcommand{\arraystretch}{1.08}
\caption{Human evaluation criteria for recommendation explanation quality (RQ3).}
\vspace{-0.2cm}
\label{tab:rq3-criteria}
\begin{tabular}{p{0.95\linewidth}}
\toprule
\textbf{Evaluation Criteria for Recommendation Explanation }\\
\midrule
\textbf{Intuitiveness:} How immediately comprehensible is the explanation? 
Does it clearly convey why the recommendation makes sense without requiring extensive interpretation?\\
\midrule
\textbf{Relatability:} How well does the explanation connect to the user's actual experiences and lifestyle context? 
Does it resonate with real-world usage scenarios?\\
\midrule
\textbf{Consistency:} How well does the explanation align with the user's demonstrated preferences and interaction history? 
Does it accurately reflect their past behavior patterns?\\
\midrule
\textbf{Persuasiveness:} How convincing is the explanation in demonstrating that the item would appeal to the user? 
Does it effectively communicate the item's value for this specific user?\\
\midrule
\textbf{Specificity:} How personalized and detailed is the explanation for this particular user-item pair? 
Does it provide concrete, individualized rationale rather than generic statements?\\
\bottomrule
\end{tabular}
\end{table}

%% file: TBL/062_humaneval-RQ4.tex
\begin{table}[t]
\centering
\small
\setlength{\tabcolsep}{4pt}
\renewcommand{\arraystretch}{1.08}
\caption{Human evaluation criteria for persona quality comparison between single-persona and multi-persona approaches (RQ4).}
\vspace{-5pt}
\label{tab:rq4-criteria}
\begin{tabular}{p{0.95\linewidth}}
\toprule
\textbf{Evaluation Criteria for Persona Quality}\\
\midrule
\textbf{Faithfulness:} Are the personas genuinely grounded in actual user reviews of this item? 
Do they reflect real opinions rather than fabricated or generic descriptions?\\
\midrule
\textbf{Clarity:} Are the personas tailored to this specific item's characteristics? 
Do they capture unique aspects rather than broad, generic user types?\\
\midrule
\textbf{Coverage:} Do the personas collectively cover diverse and meaningful reasons why different users might engage with this item? 
Are both major and minor preference dimensions represented?\\
\midrule
\textbf{Interpretability:} Are the personas logically coherent and easy to understand? 
Does each persona present a clear, well-defined user archetype?\\
\midrule
\textbf{Representativeness:} Are the personas sufficiently distinct from each other? 
Do they effectively represent different user segments rather than redundant variations?\\
\bottomrule
\end{tabular}
\vspace{-0.1cm}
\end{table}

%% file: TBL/045_RQ4_Table.tex
\begin{table}[t]
\centering
\footnotesize
\setlength{\tabcolsep}{4pt} 
\renewcommand{\arraystretch}{1.05} 
\caption{Impact of the number of generated personas per item with \bprmf in \amazon.}
\vspace{-5pt}
\resizebox{\linewidth}{!}{
\begin{tabular}{ccccccc}
\toprule
\multirow{2.5}{*}{\textbf{\# Personas}} & \multicolumn{2}{c}{\textbf{@5}} & \multicolumn{2}{c}{\textbf{@10}} & \multicolumn{2}{c}{\textbf{@20}} \\
\cmidrule(lr){2-3} \cmidrule(lr){4-5} \cmidrule(lr){6-7}
 & \textbf{HR} & \textbf{NDCG} & \textbf{HR} & \textbf{NDCG} & \textbf{HR} & \textbf{NDCG} \\
\midrule
1 & 0.0488 & 0.0301 & 0.0824 & 0.0408 & \multirow{2}{*}{$\vert$} & 0.0507 \\
3 & 0.0519 & 0.0322 & 0.0835 & 0.0423 & \multirow{2}{*}{\textbf{0.1212}} & 0.0520 \\
5 & \underline{0.0532} & \underline{0.0335} & \underline{0.0843} & \underline{0.0435} & \multirow{2}{*}{$\vert$} & \underline{0.0529} \\
7 & \textbf{0.0541} & \textbf{0.0338} & \textbf{0.0848} & \textbf{0.0437} &  & \textbf{0.0530} \\
\bottomrule
\end{tabular}
}
\vspace{-0.1cm}
\label{tab:persona-ablation}
\end{table}

%% file: TEX/050conclusion_WWW.tex
This work introduced \proposed, a novel recommendation framework that shifts online LLM reasoning to the offline by constructing diverse review-grounded item personas. 
By reframing recommendation as a user-persona alignment task trained on LLM-derived supervision, it enables real-time inference with lightweight encoder scoring while naturally providing interpretable explanations.
Extensive experiments show that \proposed consistently outperforms state-of-the-art LLM-based rerankers, achieving robust gains for cold users and tail items while reducing inference latency by orders of magnitude.
Overall, \proposed offers a practical and scalable solution bridging efficiency, accuracy, and interpretability, and paves the way for extending persona-based alignment to richer user feedback and hybrid reasoning framework that combines offline knowledge construction with adaptive online refinement.
%